\documentclass{elsart}
\usepackage{amssymb}
\usepackage{epsf}
\begin{document}

\begin{frontmatter}

\title{Manipulation of ultracold atomic mixtures using microwave techniques}
\author{D. Ciampini}, \author{E. Courtade}, \author{C. Sias}, \author{D. Cossart}, \author{G. Carelli}, \author{F. Mango}, \author{O. Morsch}, \author{E. Arimondo}
\address{INFM, Dipartimento di Fisica E.Fermi, Universit\`{a}
di Pisa, Largo Pontecorvo 3, I-56127 Pisa,Italy}

\date{\today}

\begin{abstract}
We  used microwave radiation to evaporatively cool a mixture of of
$^{133}$Cs and $^{87}$Rb atoms in a magnetic trap. A mixture
composed of an equal number (around 10$^{4}$) of Rb and Cs atoms
in their doubly polarized states at ultracold temperatures was
prepared. We also used microwaves to selectively evaporate atoms
in different Zeeman states.
\end{abstract}
\end{frontmatter}
\maketitle

\section{Introduction}
Laser cooling techniques, combined with evaporative cooling in
magnetic traps, allow the Bose-Einstein condensation (BEC) of a
dilute atomic gas \cite{ketterle99}. Typically, the atoms are
trapped in conservative magnetic potentials and ultracold
temperatures are reached through evaporative cooling using, for
instance, radiofrequency (rf), microwave (mw) or circle-of death
techniques. The evaporation methods have different applicability
and efficiency, which is more evident when dealing with
combinations of ultra-cold atoms.  A few combinations of
ultra-cold atoms have been studied in conservative traps, among
them
Li-Cs~\cite{mosk01}, K-Rb~\cite{micro,ferrari02,jin04}, and Na-Li \cite{hadzi02,stan04}.\\
\indent In experiments on ultracold atoms and atomic mixtures the
use of mw radiation in evaporative cooling has been explored by a
number of groups and has become a commonly used technique in the
field. For example, controlled state selective evaporation of a
single species was applied in \cite{chin}. Selective evaporative
cooling of a single species in a two-species magnetic trap was
used in Ref. \cite{ferrari02,truscott,ketterle02}, for the
sympathetic cooling of $^{6}$Li by $^{7}$Li in \cite{schreck} and
for sympathetic cooling of $^6$Li by $^{23}$Na in \cite{hadzi02}.
We have applied mw evaporation to a mixture of ultra-cold
$^{87}\mathrm{Rb}$ and $^{133}\mathrm{Cs}$ atoms in a magnetic
trap in order to explore sympathetic cooling and the collisional
rethermalization between the two species, as reported in Ref.
\cite{anderlini05}. With the aim of producing a very cold Cs
sample, we explored the temperature limits associated with
circle-of-death, rf and mw evaporation. This detailed exploration
is the topic of the present work.\\
\indent Our best approach for evaporating the Rb-Cs mixture was
the sequential application of different evaporative cooling
processes: an initial stage of circle-of-death evaporation active
on both species, followed by direct mw evaporative cooling of Cs,
and finally rf cooling of Rb that led to sympathetic cooling of Cs
through collisional energy exchange with Rb atoms. Applying the
usual sequence of circle-of-death followed by radio-frequency
evaporation for achieving BEC in a TOP trap, we verified that, to
a good approximation, circle-of-death evaporation is equally
efficient on Cs and Rb atoms.  We discovered that owing to the
large Cs-Rb interspecies scattering length $a_{Rb-Cs}\simeq 590
a_{0}$\cite{anderlini05} sympathetic cooling of Cs atoms by collisions
with Rb is a very effective process.  However, the Rb atom population
is substantially depleted before very low Cs temperatures are achieved.
Therefore, in order to reach very low Cs temperatures, we introduced an
intermediate stage of mw evaporation. This sequence of different
evaporation techniques was optimised in order to allow us to
measure the Cs-Rb relative scattering length at a
temperature of 6 $\mu$K, as we reported in ~\cite{anderlini05}.
In the case where rf evaporation of Rb was applied without the
intermediate step of microwave evaporation, the lowest temperature
achieved for performing the scattering length measurement was at least two
times larger.\\
\indent While characterizing the Rb-Cs ultracold mixture we faced
the presence of ultracold atoms in Zeeman sublevels different from
the desired doubly polarized state. For several degenerate gas
investigations,  an incomplete atomic magnetization represents a
great difficulty. Therefore, we employed a mw irradiation
technique to selectively evaporate Rb or Cs atoms in specific
Zeeman sublevels\cite{note1}. While rf evaporation drives
transitions between all the Zeeman sublevels of a hyperfine state,
the applied mw evaporation acts only on a given Zeeman sublevel.
We report magnetization measurements for the Rb and Cs atoms
before and after selective mw evaporation. This magnetization
purification process could be useful for exploring the
dependence of the scattering length on the Zeeman state.\\
\indent This manuscript is organized as follows: Section II
describes the experimental setup used for the preparation of the
cold Rb-Cs mixture,  the generation of mw radiation and its
delivery to the atomic sample and mw spectroscopy within the
magnetic trap. The evaporative cooling of the atomic mixture by mw
radiation is described in Section III, while the state-selective
removal of atoms is in Section IV. Section V presents some
conclusions.

\section{Experimental setup and techniques}
\subsection{Optical components}
\begin{figure}[ht]
\centering\begin{center}\mbox{\epsfxsize 3.0 in
\epsfbox{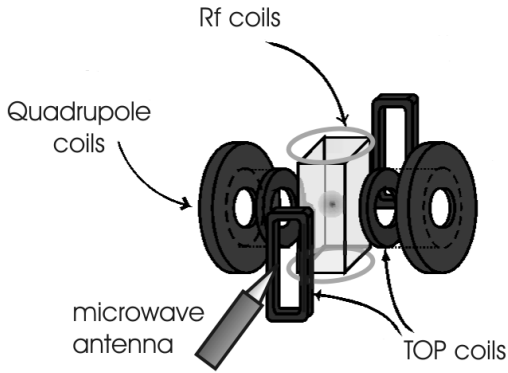}} \caption{Schematic of the coils for the
magnetic trap. The quadrupole axis is along the horizontal
direction while the bias field, produced by the TOP coils, rotates
in the horizontal plane. The rf coils and the mw antenna
are visible.}\label{schema}
\end{center}
\end{figure}

We used a double-chamber vacuum system with a 2D collection MOT
and a six-beam MOT~\cite{muller00,anderlini05}. Once the two
species MOT was filled, after brief compressed MOT and molasses
phases the trapping beams were switched off and the atoms were
optically pumped into the $|F=2, m_F=2\rangle$ and $|F=4, m_F
=4\rangle$ doubly polarized states of Rb and Cs, respectively.
Immediately after that, the TOP magnetic trap was switched on. The
bias field $B_{0}$ of the TOP rotating at 10 KHz in the horizontal
plane was created by two pairs of coils, fed by a common function
generator through a 90 degree phase shifter. A schematic of the
coils for the TOP
trap is presented in Fig.~\ref{schema}.\\
The temperature of the atomic mixture was then lowered first by
circle-of-death evaporation and afterwards  by radiative
evaporation (rf and/or mw radiation). The atom number and
temperature were measured by flashing on a beam resonant with one
of the atomic species and recording the resulting shadow cast on a
CCD camera by the atom cloud. For a given atom, different Zeeman
sublevels can be magnetically trapped and, after switching off the
magnetic trap, all of them were imaged by the same resonant laser
flash. In order to observe their relative populations for Rb we
used a very shallow magnetic potential in order to maximize the
spatial separation along the vertical direction between the
equilibrium positions of the sublevels~\cite{muller00}, the so
called gravity sag. For Cs atoms it was not possible to spatially
distinguish the atoms in the $|F=4, m_F=4\rangle$ Zeeman sublevel
from those in the $|F=4, m_F=3\rangle$ state using the
differential gravity sag, the ratio of the magnetic moments of the
two trapped states being only $\frac{4}{3}$, compared to the value
of $2$  for Rb. Thus in order to monitor the Cs atoms in different
sublevels as separate clouds, we performed a Stern-Gerlach type
experiment (3ms in which the atoms are in the quadrupole field only)
separating the Zeeman levels in time-of-flight.\\

\subsection{Microwave source}
The mw radiation was produced by a frequency locked oscillator
system composed of a Sweep Oscillator (HP $8350$B), a Frequency
Counter (HP $5343$A) and a Source Synchronizer (HP $5344$A). The
frequency sweep and the mw switch were computer-controlled and
synchronized to the experimental cycle. The mw radiation was
amplified up to $5\,\mathrm{W}$ by a second power amplifier
(Kuhne, model KU$702$ for the C-band, 4-8 GHz, and model KU$922$
for the X-band, 8-12 GHz). The mw radiation was delivered to the
atoms by a dielectric rod antenna. In our experiment we used two
different antennas, one made of teflon, optimized to deliver
radiation at $9.2\,\mathrm{GHz}$ and one made of plexiglass, used
in the region around $6.8\,\mathrm{GHz}$. The antennas consist of
a circular-section dielectric rod directly inserted into the
circular end of a metal waveguide for the X (Cs) and C (Rb) bands.
The rod is conically shaped at both ends, with the cone lengths
equal to $6 \, \mathrm{cm}$ and $8 \, \mathrm{cm}$ for the Cs and
Rb antennas, respectively. Such linearly tapered-rod antennas have
been extensively studied, and the presence of a dielectric close
to the conductor structure profoundly modifies the performance of
the antenna~\cite{libro}.\\ \indent In a preliminary experiment we
verified the behavior of the mw antennas by monitoring the
reflection coefficient with a network analyzer over a band of
frequencies around the desired frequency.
\begin{figure}[ht]
\centering\begin{center}\mbox{\epsfxsize 3.0 in
\epsfbox{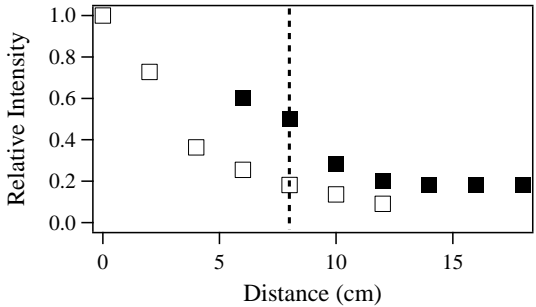}} \caption{Relative intensity of the e.m. field
measured at various distances from the end of the X-band metal
waveguide, in the presence (solid squares) and in the absence
(open squares) of the dielectric rod extension. The dashed line
marks the distance where the cold atoms are located during the
experiment.}\label{punta}
\end{center}
\end{figure}
For both antennas, the reflected power fraction never exceeded
$-10\,\mathrm{dB}$, with a modulation structure related to the
presence of objects adjacent to the antenna.  The effect of the
dielectric rod is to concentrate the mw field in a lobe in the
direction of the cone termination~\cite{Rod}, and the directivity
gain is determined primarily by the antenna length. We
experimentally found that for the same on-axis distance from the
conductor waveguide, the field intensity was amplified by a factor
2.4 when the dielectric rod was inserted (see fig.~\ref{punta}).
We also noticed a large reflection of the mw radiation due to
metallic objects located in
front of the antenna.\\

\subsection {Microwave spectroscopy}
\begin{figure}[ht]
\centering\begin{center}\mbox{\epsfxsize 3.5 in
\epsfbox{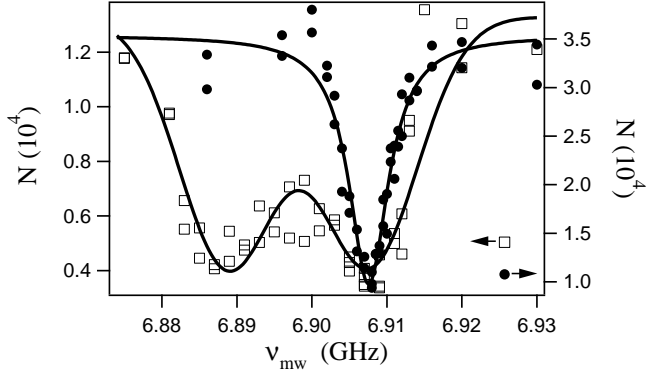}} \caption{Number of Rb atoms after application
of mw radiation for 10 s in the TOP trap as a function of the mw
frequency for two different choices of the relative amplitude of
the oscillating fields of the TOP trap, leading to a 30 percent
ellipticity (open squares, left axis) and to a circular  rotating
bias field, compensated in ellipticity and phase as described in
the text (full circles, right axis). The full lines are double
Gaussian and Lorentzian fits, respectively, to the data
sets.}\label{righe}
\end{center}
\end{figure}

We performed mw precision spectroscopy of the trapped Rb and Cs
atoms by inducing transitions between their hyperfine levels
$|F=I+1/2, m_{F}=I+1/2\rangle$ and $|F=I-1/2, m_{F}=I-1/2\rangle$.
As pointed out previously~\cite{matthews,harber,treutlein}, for a
cloud of magnetically trapped atoms the inhomogeneity due to the
energy level shifts broadens the transition frequency and limits
the attainable precision. Furthermore, in a TOP trap the presence
of the time-varying field of the trap introduces a time-varying
detuning that greatly complicates the interaction between trapped
atoms and mw radiation~\cite{martin}. Our mw spectroscopical
investigation within the TOP trap for an interaction time long
compared to the rotating field period demonstrated  a strong
dependence of the mw transition linewidth on the temporal
variation of the modulus of the rotating bias field. The analysis
of the condensate micromotion in the TOP \cite{mullerprl00} shows
that the atoms follow an orbit corresponding to a constant total
modulus of the magnetic field if the amplitude $B_{0}$ of the bias
field is constant during the horizontal rotation. On the other
hand, the total magnetic field experienced by the atoms during
their motion contains components varying at the rotating field and
its harmonics  in presence of an elliptical rotating bias field.
Such an elliptical bias field appears, for instance, if the two
linearly oscillating magnetic fields whose superposition produces
the rotating bias field are not very precisely matched in
amplitude and phase. This behavior is confirmed by the data of
Fig.\ref{righe} for the number of Rb atoms remaining in the
magnetic trap after 10 s application of 5W mw radiation whose
frequency was scanned over the resonant value. For a 30 percent
ellipticity of the bias field we observed a well-defined double
peak structure for the mw transition (open squares). When the
ellipticity of the bias field was compensated by adjusting the
relative value of the magnetic field amplitudes in the two pairs
of TOP coils and the phase, the two peaks coincided (solid
circles). This method is very sensitive: operating at our maximum
amplitude of the bias field, a phase mismatch of 5 degrees between
the two oscillating fields and an ellipticity of 10$\%$ broadens
the mw transition by $\sim$ 4 MHz. We found this fine adjustment
procedure of the bias field based on the linewidth of the mw
transition to be more sensitive than the procedure based on the
optimization of the circle-of-death evaporation efficiency.
Moreover, we performed mw spectroscopy within the TOP trap for an
interaction time short compared to the period of rotation of the
TOP field, and finally performing spectroscopy in time of flight
in the presence of a rotating bias field or a weak homogeneous
magnetic field. In this sequence of experiments we eliminated the
residual inhomoegenous broadening and measured a decreasing
linewidth. The final limiting linewidth at full mw power, 5 W (the
power injected into the waveguide), was around 100 kHz, in good
agreement with the theoretical value obtained by estimating the
amplitude of the mw magnetic field ($\sim$70 mG) at the position
of the atoms from the power inserted into the waveguide and the
enhancement factor due to the teflon cone (as seen in Fig.
\ref{punta}).

\section{Evaporation procedure}
\subsection{Circle of death}
In a TOP trap the temperature of the atomic mixture can be lowered
through circle-of-death evaporative cooling, defined by the
rotating zero of field created by the (static) quadrupole and the
(rotating) bias field. By continuously reducing the strength of
the rotating bias field the circle-of-death shrinks and atoms from
the high energy tail of the distribution are progressively removed
from the sample. In the double polarized Zeeman states Cs and Rb
atoms have the same magnetic moment. Therefore, at the same
temperature they have the same spatial extension in a magnetic
trap, and  circle-of-death evaporation is simultaneously applied
to both species. At low temperatures and high atomic density,
circle-of-death evaporation becomes inefficient because it
requires a continuous increase of the trapping frequencies leading
to an increased atomic density, resulting in large three-body
losses. At the end of our circle-of-death evaporation phase, Rb
and Cs atoms were in thermal equilibrium at $\sim
15\,\mu\mathrm{K}$.
\subsection{Radiative evaporation: radiofrequency}
Radiative evaporation  uses an e.m.
radiation field to transfer atoms from a trapped to an untrapped
state in an energy-selective way. The advantages of
radiative evaporation (rf and mw) are that the
magnetic potential does not have to be modified to sustain the
evaporation since the escape rate is precisely controlled by
the amplitude and the frequency of the applied radiation.\\
\indent Rf induced evaporation between $|F, m_F =F\rangle$
and $|F, m_F =F-1\rangle$ states exploits spin-flips of atoms on a
resonant energy shell defined by
\begin{equation}
g_{F} \mu_B B(\vec{x}) = \hbar \omega_{rf} \label{eq:EvapCond}
\end{equation}
where  $B(\vec{x})$ is the modulus of the instantaneous local
magnetic field defining the cut energy of the shell, $\mu_B$ is
the Bohr magneton, $g_{F}=2/(2I+1)$ is the Land\'{e} factor, with
$I=3/2$ for $^{87}$Rb and $I=7/2$ for $^{133}$Cs, and
$\omega_{rf}$ is the rf field frequency. Notice that
Eq.~\ref{eq:EvapCond},  and the  following one for the mw
evaporation, is derived in the limit of weak Zeeman splitting. For
$F>1$ the rf transition defined by Eq. (1) transfers atoms into a
different Zeeman state that is still magnetically trapped.
However, because of power broadening the rf radiation induces a
chain of transitions between different Zeeman levels starting from
trapped states to untrapped ones leading to a loss of atoms from
the magnetic trap. By ramping down the frequency $\omega_{rf}$,
the radius of the surface volume where the resonance condition is
fulfilled shrinks, leading to an effective forced
evaporation~\cite{ketterle99}.\\
\indent Eq.~\ref{eq:EvapCond} specifies that rf evaporation
depends on the Zeeman-sublevel spacing. When rf radiation is
applied to a mixture of atoms, the cut  energy of the atoms to be
evaporated is different. In a Rb-Cs mixture,  owing to the
difference in the Land\'e factor, the Cs cut energy  is half that
for Rb atoms. Thus, when the most energetic Cs atoms are removed
by the rf radiation, cold Rb atoms at the bottom of the potential
are also removed.  On the contrary, cooling the Rb atoms with rf
has no direct effect on the Cs temperature, because the rf field
is not resonant with the trapped Cs atoms.

\begin{figure}[ht]
\centering\begin{center}\mbox{\epsfxsize 3.0 in
\epsfbox{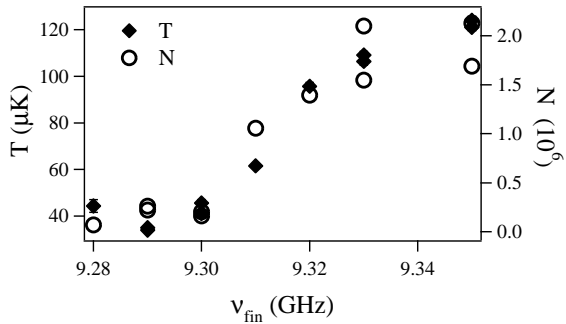}} \caption{Temperature (full diamonds)
and number of atoms (open circles) of the Cs cloud as a function
of the final value of the mw frequency. Initial frequency
$9.40\,\mathrm{GHz}$, trap bottom $9.271\,\mathrm{GHz}$, duration
of the evaporation ramp $5\,\mathrm{s}$ and mw power  $3\,
\mathrm{W}$. These data were from single experimental runs and the
uncertainties due to the fit of the single images were smaller
than the size of the data points.}\label{Fig3}
\end{center}
\end{figure}

\subsection{Radiative mw evaporation}
\begin{figure}[ht]
\centering\begin{center}\mbox{\epsfxsize 3.0 in
\epsfbox{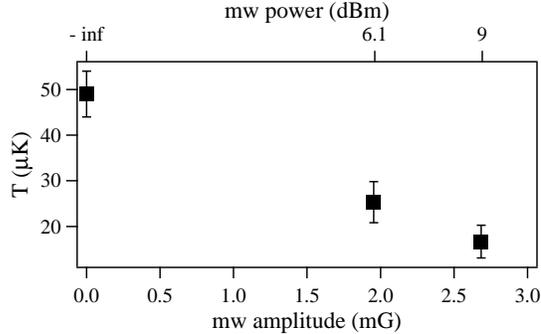}} \caption{Cs temperature as a function of
the mw power (in the upper horizontal scale), and of the mw field
amplitude at the atom position (in the lower scale),  for a given
evaporation ramp (from $9.3\,\mathrm{GHz}$ to $9.22\,\mathrm{GHz}$,
trap bottom $9.215\,\mathrm{GHz}$) with $5\,\mathrm{s}$ duration.
The data uncertainty  is due to statistical averaging over several
experimental runs.}\label{Fig4}
\end{center}
\end{figure}

For a mixture of atomic species mw evaporative cooling represents
an efficient alternative to rf evaporative cooling. Mw evaporation
uses transitions between Zeeman sublevels belonging to
\textit{different} hyperfine levels of the ground states. For a
transition between the hyperfine states $|F, m_F =F\rangle$ and
$|F-1, m_F =F-1\rangle$ of an alkali atom, the resonance condition
is
\begin{equation}
\hbar \omega_{hf} + g_{F} (2m_{F}-1) \mu_B B(\vec{x}) = \hbar
\omega_{mw} \label{eq:EvapCondmw}
\end{equation}
where $\omega_{hf}/2 \pi$ is the hyperfine splitting ($\sim
9.2\,\mathrm{GHz}$ for Cs, $\sim 6.8\,\mathrm{GHz}$ for Rb), and
$\omega_{mw}/2 \pi$ the mw frequency. The previous equation
indicates that the mw resonance conditions for Cs and Rb are
always very different and the two evaporation processes are
independent.  \\
\indent Fig.~\ref{Fig3} shows the temperature and atom number of
the Cs cloud at the end of mw evaporation in the absence of  Rb
atoms in the magnetic trap. During the evaporative cooling the
mean temperature of the Cs cloud decreased in direct proportion to
the number of atoms. The cloud was irradiated by a mw frequency
ramp with a fixed starting frequency and varying final frequency.
The evaporation took place after a magnetic compression phase and
a circle-of-death cooling stage. The data points at the far right
describe the initial condition before the application of the mw
radiation. The dependence of the final temperature of the Cs cloud
on the mw intensity is shown in Fig.~\ref{Fig4}.
\indent In our experiments we generally observed a lower
efficiency for the mw evaporation compared to rf evaporation. Due
to imperfect optical pumping (and, possibly, other depolarizing
processes during the evaporation cycle), both the Rb and Cs cold
clouds had admixtures of atoms in other Zeeman sublevels, as
discussed in the following Section. While the radio-frequency
radiation, inducing a chain of transitions between the equally
spaced Zeeman sublevels, evaporates atoms from all of them, the mw
transitions of Eq. \ref{eq:EvapCondmw} are well separated in
frequency, at $30-40 \mathrm{G}$ trap field and in the absence of
saturation broadening. Thus, the mw field is resonant with one
Zeeman state only, leaving the residual population of the other
sublevels uncooled by the mw radiation. This residual population
is eventually cooled through sympathetic cooling collisions with
the atoms in the other Zeeman state, but the overall efficiency of
the mw evaporation is reduced. From the data of Fig. \ref{Fig3}
for Cs, assuming an atomic occupation of the single Zeeman
sublevel $|F=4, m_F =4\rangle$,  we derived that under mw
evaporation the increase of the phase space density followed a
$N^{-2}$ law with $N$ the atom number, a law similar to that
required for the runaway regime reached by the rf evaporation
\cite{ketterle99}. However, the occupation of the $|F=4, m_F
=3\rangle$ sublevel for the data in Fig. \ref{Fig3} is around
30-40$\%$, as extracted from the Stern-Gerlach type experiments
performed with ultracold Cs atoms (see Section IV), and that
occupation greatly decreases the effective efficiency of the
phase-space compression by the mw evaporation.

\subsection{Sympathetic cooling}
After the preparative stages of circle-of-death and mw evaporation
we verified that when performing rf evaporation on the Rb atoms,
the measured temperature of the Cs atoms exactly followed the Rb
temperature down to a few $\mu \mathrm{K}$, indicating that
sympathetic cooling was taking place, as long as the remaining
number of Rb atoms, compared to the Cs atom number, was sufficient
to sustain the thermalization. By applying mw techniques we were
able to prepare a mixture of $4\times 10^4$ Rb atoms in the $|F=2,
m_F=2\rangle$ state and $10^4$ Cs atoms in the $|F=4, m_F
=4\rangle$ state at about $6\, \mu\mathrm{K}$ (mean trapping
frequency $70\, \mathrm{Hz}$). This temperature is of the same
order of magnitude as that reported by the ENS group~\cite{ENS}
for Cs in the same atomic state in their early search for
condensation of Cs.  From our data we extrapolated that for Cs a temperature
around $3 \mu$k could be achieved if all of the Rb was evaporated.
A detailed model of the sympathetic cooling
phase was developed in ref.~\cite{anderlini05} in order to derive
the Rb-Cs interspecies scattering length.

\section{State selective evaporation}
\begin{figure}[ht]
\centering\begin{center}\mbox{\epsfxsize 3.0 in
\epsfbox{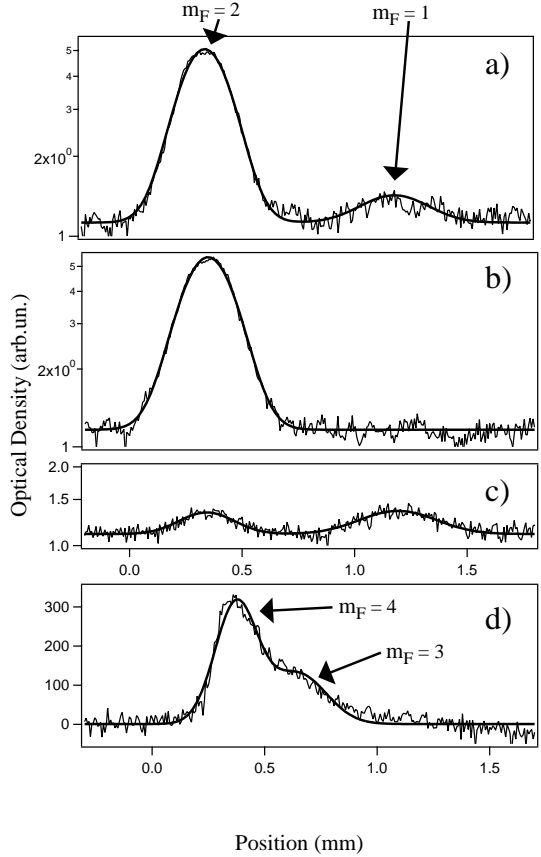}} \caption{Rb integrated absorption
profile in free fall (a) before and (b), (c) after selective mw
evaporation, as in the text. The two peaks correspond to the
$|F=2, m_F=2\rangle$ and $|F=2, m_F=1\rangle$ states experiencing
a different gravitational sag for a mean trap frequency of $25.7
\, \mathrm{Hz}$, with the atomic position measured downwards from
the quadrupole symmetry center. In (b), (c) and (d) mw evaporative
cooling produced a purification of Rb magnetization, as in the
text. In (d), integrated absorption profile of Cs atoms, detected
though a Stern-Gerlach type experiment, after selective removal of
atoms in the $|F=4, m_F=3\rangle$ state, with the procedure
explained in the text. Note that for better visibility we chose in
(a), (b) and (c) a logarithmic vertical scale, while in (d) the
vertical scale is linear.} \label{Fig5}
\end{center}
\end{figure}

Another application of the state selectivity provided by the mw
evaporation technique is state purification. Since at the end of
the circle-of-death evaporation phase we measured a percentage of
around 80 percent of Rb atoms in the desired doubly polarized
state $|F=2, m_F=2\rangle$, with around 20 percent in the $|F=2,
m_F=1\rangle$ sublevel, we used the mw radiation to eliminate the
populations in the sublevels other than the desired doubly
polarized state. This is not achievable using rf radiation since,
owing to saturation broadening, the resonance condition of
Eq.~\ref{eq:EvapCond} is equally fulfilled for atoms in all
trapped Zeeman states. The results of the state selective
evaporation in a Rb atomic sample is shown in Fig.~\ref{Fig5}. The
two peaks in the upper record correspond to the absorption profile
of 0.8 $\mu$K Rb atoms in the different Zeeman states, spatially
separated along the vertical direction because of the differential
sag. In (b) atoms in the $|F=2, m_F=1\rangle$ state have been
reduced from 10 percent to an undetectable level, with no loss of
atoms from the $|F=2, m_F=2\rangle$ state, when 100 mW  mw
radiation at 6.858 GHz resonant with the the hyperfine transition
at the bottom of the magnetic trap was applied for
$10\,\mathrm{s}$.  Even if the mw radiation was not resonant with
the whole sample because of inhomogeneous broadening,
rethermalizing collisions allowed the evaporation to act on the
whole hyperfine level occupation. In (c) mw radiation at 6.904 GHz
was applied to the $|F=2, m_F=2\rangle$ hyperfine level with a
residual six percent final occupation in that state.\\
\indent For Cs atoms, performing a Stern-Gerlach type experiment
separating the Zeeman levels in time-of-flight at the end of the
circle-of-death evaporation phase we detected relative populations
of Cs atoms in the $|F=4, m_F=4\rangle$ Zeeman sublevel and in
$|F=4, m_F=3\rangle$ of 60 percent  and 40 percent, respectively.
The percentage of atoms in the unwanted Zeeman sublevel was higher
for Cs atoms than for Rb atoms. In Fig. \ref{Fig5} (d) the density
profile of a Cs cloud is shown after the application of 5 s of mw
radiation at 9.250 GHz resonant with  the $|F=4, m_F=3\rangle$
$\rightarrow$ $|F=3, m_F=2\rangle$ transition at the bottom of the
trap, followed by a Stern-Gerlach phase, with a remaining fraction
of $m_F=3$ atoms of 30$\%$. The number of Cs atoms before the
application of the mw was around $10^4$ and their temperature
$\sim$4 $\mu$K at 51 Hz mean trapping frequency. By applying a 10 s
mw evaporation phase, similar to that applied to the Rb atoms, we
were able to reduce the fraction of atoms in the $|F=4,
m_F=3\rangle$ state to about 20 percent. The different Zeeman
occupations for Rb and Cs that we found to start with in the
magnetic trap and the different efficiency of the selective mw
removal of atoms in the Rb and Cs case are compatible with the rate
of density-dependent inelastic collisions of Cs atoms leading to a
change of the $m_F$ quantum number, as reported by~\cite{Foot}. From
our data taken at a Cs density of $8.\times 10^{10}$ cm$^{-3}$ we
infer an inelastic rate coefficient around $1.\times 10^{-12}$
cm$^{3}$s$^{-1}$. This value is in agreement with the losses
reported in the experiment of ref. \cite{Foot} performed at a
different magnetic field and a higher temperature.
\section{Conclusions}
In a two-species experiment, mw techniques are a very useful
method to perform radiative evaporation, in particular when
combined with other cooling techniques. Mw evaporation is very
flexible with respect to the choice of the magnetic trap confining
the atoms, because even for a large change of the local magnetic
field determining the mw resonance,  the shift of the mw resonance
frequency remains within the emission bandwidth of the mw source.
We showed that both Rb and Cs atoms can be efficiently evaporated
using mw radiation. For a Rb-Cs mixture the insertion of a Cs mw
evaporation stage allowed us to reach temperatures in the 15
$\mu$K range for both species. A similar temperature could be
reached  through sympathetic cooling only wasting approximatively
50 percent of the Rb atoms. The combination of circle-of-death, mw
and rf evaporative cooling allowed us to perform collisional
studies of the Rb-Cs mixture and to derive the value of the
interspecies scattering length, around 590
a$_{0}$\cite{anderlini05}. As the main difference between rf and
mw evaporation, owing to the presence of multiphoton transitions
the rf field depletes all Zeeman sublevels of a target hyperfine
state. By contrast, the mw evaporation is Zeeman sublevel
selective. In fact we have made use of this selectivity associated
with the mw radiation to remove either Rb or Cs atoms from
specific Zeeman sublevels, manipulating and purifying the
magnetization of our Rb-Cs mixture. Making use of the differential
gravitational sag or of the Stern-Gerlach separation we verified
the efficiency of the mw selective evaporation. On the basis  of
precise simulations for the mw or rf evaporation, a quantitative
comparison between the efficiencies reached in rf and mw
evaporation could be performed. Furthermore additional information
on the elastic, and inelastic, collisional properties of the
ultracold mixture could be derived. However owing to the
complexity of atomic Zeeman/hyperfine level structure such a
simulation represents a very difficult task. Finally, we used mw
spectroscopy to precisely calibrate the ellipticity of the
rotating bias field of a TOP trap. All these applications of the
mw evaporation are useful for the preparation of an atomic mixture
in precisely controlled conditions.


\section{Acknowledgements}
We thank J. Reichel for preliminary discussions on the microwave
equipment, A. Piombini for the network analyzer measurements, M.
Giordano for the loan of microwave equipment, and A. Alberti and
G. Tumino for help in the early stages of the experiment. This
research was supported by the INFM (PRA Photonmatter), and by the
EU Network Cold Quantum Gases.



\begin{thebibliography}{99}
\bibitem{note1} In \cite{jinPRL99} the relative occupation of the
$^{40}$K Zeeman levels was controlled by mw evaporation.
\end{thebibliography}
\end{document}